\documentstyle[12pt]{article}

\begin{document} 
\title{An interpretation of the black-body radiation
}
\author{Shun-ichiro Koh\\ 
        Physics Division, Faculty of Education, Kochi University  \\
        Akebono-cho, 2-5-1, Kochi, 780, Japan
         \thanks{ e-mail address: koh@cc.kochi-u.ac.jp } \date{}
	е	
}
\maketitle
   \begin{abstract}
    The black-body radiation is reinterpreted in terms of the 
   photon's many-body wave functions in analogy with the condensed matter physics.
   This interpretation has implications on the wave-particle duality, 
   and on the difference between the photon and the matter wave.
   \end{abstract}е
   
              
\newpage
The black-body radiation, a photon gas coupled 
with the matter in the thermal equilibrium, was historically a starting 
point for the birth of the quantum physics.  The low-frequency region of 
the Planck distribution at high temperature agrees with the result 
by the classical electromagnetism, and its high-frequency region 
at low temperature represents the particle nature of 
the light (the photon), which characterizes the black-body radiation 
as a phenomenon representing the wave-particle duality \cite{wol}. 
The wave-particle duality is now a conceptual 
basis of the quantum physics, but it continues to be a problematic idea 
which is beyond our imagination. As a result, the light is often 
explained in such a way that it changes its nature according to the 
apparatus it is interacting with: The light is a wave when passing 
through a pair of slits, but it is a stream of photons when it strikes a 
detector. 

Normally the  Planck distribution is derived on the picture of the independent 
particles obeying the Bose statistics. The statistical 
mechanics of the black-body radiation is sometimes regarded as a subject
 which has been already understood. 
This paper points out that, when the Planck distribution is reinterpreted  
using the concept of {\it the many-body wave function \/} 
(in analogy with the condensed matter physics),
there still exists some points to be considered. The typical  
Bose particle in the condensed matter physics is the  helium-4.
 In the study of the helium-4, Feynman proposed an interesting method of 
 introducing the Bose statistics which emphasized the role of the 
 many-body wave function \cite{fey}.
 His method gave us the same result as the conventional one for 
 the helium. But when it is applied to the photon,
 it discloses the significant nature of photon which has not been 
 stressed until now. 
 This paper applies his method to the photon, and discusses its implication. 

Consider $N$ photons having a following partition function in a box,
\begin{equation}
	Z=\sum_{{p}}<p_1,\ldots p_N |\exp(-\beta H) |p_1,\ldots p_N>еее.
	\label{е}
\end{equation}е
To grasp the deep meaning of the quantization, it is useful to begin with 
the classical picture, and to focus on  where the quantum feature 
is introduced. If the photon is treated as if it is a massless classical particle 
with an energy $\hbar cp$, the partition function for a single photon 
 has such a form as, 
\begin{equation}
	Z_1=\sum_{p}<p|\exp(-\beta \hbar cp)|p>е.ее
	\label{е}
\end{equation}е
The quantum nature of the photon appears in the $N$-photon system obeying
 the Bose statistics. The permutation symmetry is imposed, but its 
 appearance is affected by the thermal disturbance.   
 We assume that,  among the $N$ photons,        
$s$ photons having the same momentum and polarization form a coherent many-body 
wave function, in which the permutation symmetry is satisfied in such a 
way that an interchanging of the $s$ photons leaves the 
wave function unaltered. 

Here, the phrase ``wave function consisting of $s$ photons'' must be used 
carefully. The wave function gives information 
about the possibility of one photon in a particular place and not the probable 
number of photons in that place. (As Dirac stated, each photon interferes 
only with itself \cite {dir}.) Hence, each photon has its own wave 
function spreading out in the coordinate space, and in some cases it 
reaches to the macroscopic distance. Here, the number $s$ does not 
reflect the spatial size of the many-body wave function, but it reflects 
the number of photons connected by the Bose statistics. 
(In this paper, we call the many-body wave function consisting of many photons 
 a macroscopic wave function with this meaning.)
 
From now, we show that the photon's many-body wave function gives us a 
natural interpretation of the black-body radiation.
In view of $\exp(-\beta \hbar cp)$ in Eq.(2), the many-body wave function 
consisting of $s$ photons with the same momentum 
will contain  $\exp(-\beta \hbar cps)$ in $Z$ (Eq.(1)). The integral of
$\exp(-\beta \hbar cps)$ over the length and orientation of $p$ such as,
\begin{equation}
	f_s=2\int_{0}^{\inftyе}\exp(-\beta \hbar cps)4\pi p^2\frac{dp}{(2\pi)^3е}
	   =\frac{2}{\pi ^2е}\left(\frac{k_BT}{\hbar cе}\right)^3\frac{1}{s^3е}  ,еее
	\label{е}
\end{equation}е
is a quantity representing the many-body wave function consisting 
of $s$ photons with the same momentum and polarization, 
which will serve as an element for interpreting $Z$ in terms of
the many-body wave function. (A factor 2 in front of 
the integral comes from the photon's helicity states.)

The partition function derived from the Planck distribution has a following form,
\begin{equation}
	Z=\exp\left(\frac{(k_BT)^3}{3\pi ^2(\hbar c)^3е}
	        \int_{0}^{\inftyе}\frac{p^3dp}{e^p-1е}еее\right)е.ее
	\label{е}
\end{equation}е
Using the Riemann zeta function,
\begin{equation}
	\zeta(r)=\int_{0}^{\inftyе}\frac{x^{r-1}dx}{e^x-1е}
	   =\Gamma (r)\sum_{s=1}\frac{1}{s^rе}еее   ,
	\label{е}
\end{equation}е
$Z$ has a following form,
\begin{equation}
	Z=\exp\left[\frac{2}{\pi ^2е}\left(\frac{k_BT}{\hbar cе}\right)^3
	         \sum_{s=1}\frac{1}{s^4е}\right]еее.
	\label{е}
\end{equation}е
Expanding the exponential function and using the explicit form of $f_s$ (Eq.(3)),
one obtains,
\begin{equation}
	Z=\prod_{s=1}^{\infty е}е\sum_{\xi _s=0}^{\infty е}
	            \frac{1}{\xi _s!е}\left(\frac{f_s}{sе}\right)^{\xi _s}ееее.ее
	\label{е}
\end{equation}е
With a slight modification such as,
\begin{equation}
		Z=\frac{1}{N!е}ее\sum_{\{ \xi _s\} }
	           \left(\frac{N!}{\prod_{s=1}\xi _s!s^{\xi _s}е}\right)е
	             f_1^{\xi _1}f_2^{\xi _2}\cdots  f_s^{\xi _s}\cdots ееее,ее
	\label{е}
\end{equation}е
this expression allows a following combinatorial interpretation:
 In the black-body 
radiation consisting of $N$ photons, each photon belongs to one of the many-body 
wave function $f_s$ consisting of $s$ photons. These wave functions  appear 
$\xi _s$ times in the black-body radiation, a distribution of which is
 $\{ \xi _1,\ldots, \xi _s, \ldots \}$ being subject to $\Sigma _s s\xi _s=N$. 
 Since the $s$ photons are indistinguishable, the bracket in the right-hand side of 
Eq.(8) agrees with the number of ways of rearranging $N$ photons into 
the wave functions having the distribution $\{ \xi _s\}$. 

The physical meaning of this wave function is as follows. In the small 
momentum at high temperature, the Planck distribution agrees with the 
energy spectrum obtained when we regard the black-body radiation as the 
classical electromagnetic wave. In view of Eq.(3), 
it is the very same small $p$ and small $\beta$ in the exponent 
that determines $f_s$  for a large $s$ in the integrand. This suggests 
that the $f_s$  for the large $s$ is related to the object of the
classical physics, or more precisely, the expectation value of 
the field operator in the many-body wave function, 
which consists of many photons having the same momentum 
and polarization, corresponds to the classical electromagnetic wave. 
 When extrapolating this 
view to the small $s$, we are led to an image of the black-body 
radiation as {\it  a random assembly of the small packets of the 
electromagnetic wave, the photon number of which is distributed
 proportionally to $1/s^3$ \/}
(Eq.(3)). At a smallest limit of the packet, we find a
photon there. Conversely, in the course of a scaling up process 
($s\rightarrow \infty $), the wave function loses
its quantum nature, making its expectation value the classical wave.

To understand the photon, it is useful to compare it with the matter wave 
having an energy $p^2/2m$. Using this energy in 
Eq.(3), one obtains,
\begin{equation}
	f'_s=\int_{0}^{\inftyе}\exp(-\frac{\beta p^2s}{2m}е) 4\pi p^2\frac{dp}{(2\pi)^3е}
	   =\left(\frac{mk_BT}{2\pi е}\right)^{1.5}\frac{1}{s^{1.5}е}    ,ееееееее
	\label{е}
\end{equation}е
Feynman showed that a substitution of $f'_s$ to Eq.(7) reproduces the 
grand partition function of the Bose distribution \cite{fey}\cite{fs}.
In view of Eq.(3) and (9), one find that $f_s$ is a more rapidly decreasing 
function of $s$ than $f'_s$, meaning that as $s\rightarrow \infty $ the 
contribution of the electromagnetic wave consisting of many photons 
falls more rapidly than that of the matter wave consisting of the same 
number of particles.

 At first sight, this result seems to be contradictory to our naive 
expectation that $f_s$ is larger than $f'_s$ for a large $s$,  
because the classical electromagnetic wave propagates 
macroscopically. But it has a following obvious reason.
Generally, in the energy spectrum of the system,  
 a component which mainly determines the statistical properties of 
the system is the small-momentum and low-frequency one. 
In the Boltzmann factor of Eq.(3) and (9),  
 $\hbar cp$ is larger than $p^2/2m$ for a small $p$.
 Hence, the  electromagnetic wave consisting of $s$ photons gives
  a smaller contribution $\exp(-\beta \hbar cps)$ to $Z$ than
 the matter wave consisting of $s$ particles gives  $\exp(-\beta p^2s/2m)$ to $Z$.
This means that, {\it compared with the matter wave, the macroscopic coherence 
of the classical electromagnetic
 wave, which is normally observed in the non-equilibrium situation, is drastically 
 suppressed in the thermal equilibrium.  \/}
(We must note that $s$ in $f_s$ and $f'_s$ does not 
correspond to the quantity directly observed in the experiments such as 
the wave packet's length in the coordinate space \cite{ykoh}. 
Rather,  $f_s$ and $f'_s$ are manifestations of the 
both waves in the statistical-mechanical formalism.) 

The wave-particle duality appears in the fluctuation as well. As is well known,
using the Planck distribution in $<\Delta E^2>=\partial ^2logZ/\partial\beta ^2$, 
the square of the relative fluctuation of the black-body radiation energy 
in a given frequency interval $\Delta \nu$ is obtained as follows,
\begin{equation}
	\frac{<\Delta E^2>}{<E>^2е}е=\frac{h\nu}{<E>е}+
	    \frac{1}{\rho (\nu)\Delta \nu е}ее.
	\label{е}
\end{equation}е
where $\rho (\nu )$ is a density of oscillations between $\nu$ and $\nu 
+\Delta \nu$ \cite{tom}.
The first term in the right-hand side shows that, as the mean energy 
increases, the relative fluctuation decreases, suggesting an 
accumulation of the many independent and small fluctuations such as a creation 
and an annihilation of the photon. The second term  does not depend on the 
mean energy, suggesting the fluctuation of the wave energy.  In the 
random assembly of the waves, the accidental superposition of two waves 
with slightly different frequencies (beats) gives rise to such a 
fluctuation. 

From a viewpoint of the photon's many-body wave function, 
this wave-particle duality in the fluctuation can be interpreted as 
follows. Using Eq.(3) and (8) in $<\Delta E^2>=\partial ^2logZ/\partial\beta ^2$, 
one obtains,
\begin{equation}
		<\Delta E^2>=\frac{3(k_BT)^2}{Zе}\sum_{i}
		    \left[\frac{1}{N!е}ее\sum_{\{ \xi _s\} }
	           \left(\frac{N!}{\prod_{s}s^{\xi _s}е}\right)е
	             \frac{f_1^{\xi _1}}{\xi _1!е}е\cdots 
	                 \frac{f_i^{\xi _iе}}{(\xi _i-1)!}е
	                 \cdots  \frac{f_s^{\xi _s}}{\xi _s!е} \cdots ееее
	                 \right].е
	\label{е}
\end{equation}е 
In $<\Delta E^2>$, a wave function consisting of $i$ photons appears $\xi _i$ times as 
frequently as others. The fluctuation of the black-body radiation is a sum 
of independent fluctuations caused by an appearance or a disappearance of 
the many-body wave functions, whose constituents range from a single photon to a 
macroscopic number of photons. 
 When the fluctuation is caused by the wave function with a large $s$, many 
photons having the same momentum and polarization participate in it at once, 
thus being observed as a fluctuation of 
the wave. On the contrary, when the fluctuation is caused by the wave 
function with small $s$, it is observed as a particle-number fluctuation. 
In view of Eq.(11), in addition to the both extreme limits (wave or 
particle fluctuation), other fluctuations having intermediate scales exist as well.

In the statistical optics, the wave-particle duality is discussed in 
terms of the correlation tensor including the two-points correlation function of 
the electric field or the magnetic field such as,  
\begin{equation}
	<E^{-}_i(r_1,t_1)E^{+}_j(r_2,t_2)>, <B^{-}_i(r_1,t_1)B^{+}_j(r_2,t_2)>.
	\label{е}
\end{equation}е
The correlation tensor of the black-body radiation
was calculated by several authors, which shows that the range of its field 
correlation in the coordinate space is of order $\hbar c/k_BT$ \cite{bou}. 
This means that the coherence volume is of order $(\hbar c/k_BT)^3$.  
In the Planck distribution, the average density of the photons has a value such as,
\begin{equation}
	\frac{2}{\pi ^2е}\left(\frac{k_BT}{\hbar cе}\right)^3\zeta(3),
	\label{е}
\end{equation}е
meaning that the average number of the photons in the coherence 
volume $(\hbar c/k_BT)^3$ is of order unity $2\zeta(3)/\pi ^2\cong 1$,\cite{man}. 
This single photon in the coherence 
volume must belong to one of the many-body wave 
functions consisting of $s$ photons.  Equation.(3) says that its 
statistical weight is proportional to $1/s^3$. This 
suggest that the interference in the black-body radiation is a phenomenon 
caused by the wave functions consisting of only a few photons. In 
our picture of the black-body radiation, $f_s$ must play an important 
role in the coherence. Probably, the correlation tensor is not the only 
quantity measuring the coherence of the black-body radiation. The 
conventional definition of the degree of the coherence must be 
interpreted in terms of the photon's many-body wave function, which is a 
subject of future paper. 

This paper discusses the wave-particle duality in the case of the 
black-body radiation. But the former is not a concept limited to the 
thermal equilibrium, but an universal one. Hence, for other 
photon's distributions that are made by passing the black-body radiation through some filter,  
if we have an appropriate statistical factor instead of the Boltzmann factor, 
we can apply the combinatorial interpretation in terms of the photon's 
many-body wave function  to such photon's distributions as well. 

Furthermore, if we find a statistical factor appropriate for the 
non-equilibrium photon's distribution, a scope of this interpretation 
expands further. This statistical factor will open a new possibility for 
understanding the wave-particle duality. For this purpose, we must find
 a statistical factor corresponding to the apparatus which the photon interacts
  with. When the light is observed by one apparatus, $f_s$ may have a weak 
$s$-dependence,   
so that the macroscopic many-body wave function, which consists of photons with
 the same momentum and polarization, appears as frequently as the 
microscopic one, thus leading to the wave picture of the light.
 On the contrary, when 
observed by another apparatus, $f_s$ may be a rapidly decreasing function of 
$s$, so that the light seems to consist of only microscopic wave function, 
thus leading to the particle picture of the light \cite{re8}.
Only when we are able to describe the behavior of light in an unified scheme,  
instead of beginning with two completely different pictures, 
we know the true meaning of the wave-particle duality. 
 This is an open problem.

Acknowledgment

The author thanks Y.Koh for a critical comment.

е


\begin{thebibliography}{99}
\bibitem{fs} For the matter wave, $f'_s$ is derived from the particle 
	picture using the path integral formalism as in Ref.2. For the photon, however, a 
	similar derivation of $f_s$ is questionable. This fact implies that, in 
	spite of their formal resemblance in the momentum space, the photon 
	in the coordinate space can not be treated in the same manner as the matter wave.
\bibitem{re8} Ref.8 proposed the photon number in the coherence volume 
	as a criterion for judging whether the light behaves like a wave or a 
	particle. Physically, this criterion agrees with ours ($s$-dependence of 
	$f_s$), but its exact relationship is worth exploring.
\end{thebibliography}
\end{document}